\documentclass[11pt]{article}
\usepackage{calc}
\usepackage[dvips]{graphicx}
\usepackage{xcolor}
\usepackage{epsfig}
\usepackage{amsmath,amsfonts,amssymb,amsthm}
\usepackage{verbatim}
\usepackage{psfrag}
\usepackage{bm}
\usepackage{bbm}
\usepackage[square,comma,sort&compress,numbers]{natbib}
\usepackage{color}
\usepackage{xcolor}
\usepackage{slashed}
\usepackage{upgreek}

\RequirePackage[colorlinks=true,urlcolor=blue ,anchorcolor=blue,citecolor=blue
  ,filecolor=blue,linkcolor=blue,menucolor=blue ,linktocpage=true,pdfproducer=medialab ,pdfa=true
]{hyperref}

\usepackage[T1]{fontenc}

\usepackage{cancel}

\usepackage{epsf,epsfig}
\usepackage{graphics}
\def\tr{{\rm tr}}
\def\VEV#1{\left\langle #1\right\rangle}
\def\sst{\scriptscriptstyle}
\def\hf{{\textstyle{1\over2}}}

\begin{document}

%\begin{flushright}
%    {IPPP-25-XXX}
%\end{flushright}
%\vspace{0.4cm}
\begin{center}
\Large\bf\boldmath
A note on instantons, $\theta$-dependence and strong CP\\
\unboldmath
\end{center}

%\vspace{0.4cm}

\begin{center}
\large Valentin V. Khoze\\
\vskip0.3cm
{\footnotesize IPPP, Department of Physics, Durham University, Durham DH1 3LE, UK \\
valya.khoze@durham.ac.uk }
%\vskip1.4cm
\end{center}

\begin{abstract}
\noindent I review the standard instanton framework for determining the $\theta$-dependence of instanton-dominated correlation functions in QCD. I then contrast these well-established semiclassical results with the recent assertion of Refs.~\cite{Ai:2020ptm,Ai:2024cnp}  that $\theta$-phases are absent and that strong interactions preserve CP for all values of $\theta$. In all scenarios considered, the $\theta$-dependence is shown to be intrinsically non-trivial, being governed either by instantons in the weak-coupling regime or by alternative, non-perturbative considerations when the theory is strongly coupled.
\end{abstract}

\paragraph{Dilute Instanton Gas approximation.} 
Consider QCD with massive quarks and include the conventionally normalised $\theta$-term,
$-\,\frac{i \theta }{16 \pi^2}\, {\rm tr}\, F_{\mu\nu} \tilde F_{\mu\nu}$, in the Euclidean Lagrangian. 
Assume that the theory is weakly coupled so that a semiclassical treatment is reliable. [Situations in which this assumption fails are discussed at the end of this paper.] 
Our goal is to compute the partition function $Z(\theta)$ and thereby determine the vacuum energy density as a function of~$\theta$,
\begin{equation}
E(\theta)
= -\frac{\log Z(\theta)}{V_4},
\label{eq:Eth}
\end{equation}
where $V_4$ is the four-dimensional Euclidean volume.
To proceed, we further assume that the dominant contributions to the path integral for $Z(\theta)$ arise from expanding around classical field configurations consisting of $n+\bar n$ widely separated, non-interacting instantons and anti-instantons.
Summing over all possible numbers $n$ of instantons and $\bar n$ of anti-instantons leads to the standard dilute instanton gas approximation (DIGA) formula~\cite{Callan:1976je,Jackiw:1976pf,Coleman1985Aspects},
\begin{equation}
Z(\theta) =\, \sum_{n=0}^\infty \sum_{\bar n=0}^\infty \frac{1}{n! \, \bar n!}
\left(\int d^4 x_0\, K_I \, e^{-\frac{8\pi^2}{g^2} +i \theta}\right)^n
\left(\int d^4 x_0 \, K_{\bar I} \, e^{-\frac{8\pi^2}{g^2} - i \theta}\right)^{\bar n}.
\label{eq:Zth}
\end{equation}
In this expression, $x_0^\mu$ denotes the position of each individual single-instanton or single-anti-instanton configuration, and $K_{I,\bar I}$ is the corresponding instanton or anti-instanton density.\footnote{%
$K_I = \frac{\int d\mu_I}{\int d^4 x_0}$, where $d\mu_I$ is the integration measure over the one-instanton collective coordinates, including the Jacobian factors and the `t~Hooft determinants of the quadratic fluctuation operators around the instanton.}
The factors $e^{-8\pi^2/g^2 \pm i\theta}$ are the semiclassical suppression factors $e^{-S_E}$ associated with the Euclidean action of a single instanton or anti-instanton, including the contribution of the $\theta$-term.
Since all quarks are assumed to be massive, the fermion zero modes of (anti-)instantons are lifted by the Dirac masses, and therefore the factors $K$ appearing on the right-hand side of~\eqref{eq:Zth} are proportional to $\det M_f$ and do not vanish.
If the fermion masses $M_f$ are taken to be complex-valued,
\begin{equation}
K_I \,=\, |K| \, e^{i \alpha}\,, \quad K_{\bar I} \,=\, |K| \, e^{-i \alpha}\,, \quad
\alpha\,\equiv\, \log {\rm arg} \det M_f
\end{equation} 
where $|K|$ is a positive constant of  mass-dimension 4, so that 
$\int d^4 x_0 \, |K| = V_4 \, |K|$ is dimensionless.
It is also convenient to define the combined phase parameter $\bar\theta \,=\, \theta + \alpha$.

The standard (and ultimately correct) procedure within the DIGA framework is to evaluate the infinite double sum on the right-hand side of~\eqref{eq:Zth} at finite Euclidean volume $V_4$,
\begin{equation}
Z(\theta) =\, \sum_{n=0}^\infty \frac{1}{n!} \left(V_4\, |K| \, e^{-\frac{8\pi^2}{g^2} +i \bar\theta}\right)^n
\times \, 
\sum_{\bar n=0}^\infty \frac{1}{\bar n!} \left(V_4\, |K| \, e^{-\frac{8\pi^2}{g^2} -i \bar\theta}\right)^{\bar n}
=\, e^{\, 2 V_4  |K| e^{-8\pi^2/g^2} \cos \bar\theta}
\label{eq:Zth2}
\end{equation}
 and then divide by the volume the logarithm of the result, as in~\eqref{eq:Eth},
\begin{equation}
E(\theta)  =\, -\,  \frac{\log Z(\theta)}{V_4} \, =\, - 2 |K|\, e^{-8\pi^2/g^2}\, \cos \bar\theta .
\label{eq:Eth2}
\end{equation}
For non-trivial values of $\bar\theta$, CP symmetry is broken by strong interactions. A popular solution to the resulting strong CP problem is provided by promoting the CP-violating phase $\bar \theta$ to a dynamical axion field $a(x)$ via $\bar \theta(x) = a(x)/f_a$, where $f_x$ is the axion decay constant. The axion potential is then read from~\eqref{eq:Eth2},
\begin{equation}
V(a)  =\, 2 |K|\, e^{-8\pi^2/g^2}(1- \cos (a (x)/f_a)) .
\label{eq:Eth2a}
\end{equation}

We emphasise that although the dilute instanton gas expression for $Z(\theta)$
involves a summation over all possible numbers of instantons and anti-instantons, the expression for 
$\log Z(\theta)$ is determined solely by the contribution of a single instanton together with its Hermitian conjugate -- namely, a single anti-instanton. In diagrammatic language, this corresponds to the sum of two minimal \emph{connected} diagrams: the single-instanton vertex and the single-anti-instanton vertex. These diagrams are `connected' in the sense that a single instanton is an elementary semiclassical configuration whose vertex cannot be factorised into products of more basic ones.

The DIGA partition function $Z(\theta) = e^{-V_4\, E(\theta)}$ in~\eqref{eq:Zth2} does receive 
contributions from multi-instanton--multi-anti-instanton sectors, but these enter only through \emph{disconnected} diagrams. This is exactly as expected: the exponential of the sum of connected diagrams automatically generates all disconnected diagrams built from them.

\begin{enumerate}
\item[] {\tt Note} that, in principle, the dilute gas approximation can be systematically improved 
by incorporating 2-particle, 3-particle, and all higher-particle interactions.  
In statistical mechanics this procedure is known as the virial (or Mayer cluster) expansion.  
For example, the pressure $p$ of a non-dilute gas is
\begin{equation}
p = k_B T\, \frac{1}{V} \log Z(T,V,\mu),
\end{equation}
where $Z(T,V,\mu)$ is the grand canonical partition function and $V$ is the volume.  
The quantity analogous to the vacuum energy $E(\theta)\,V_4$ in~\eqref{eq:Eth2} is therefore 
$-p\, V/(k_B T)$.
The virial expansion gives
\begin{equation}
\label{eq:virial}
\frac{pV}{k_B T} 
= \log Z(T,V,\mu) 
= V \kappa \left( 1 + \kappa B_2(T) + \kappa^2 B_3(T) + \ldots \right),
\end{equation}
where $\kappa$ is the particle density (the analogue of the instanton density 
$|K|\, e^{-8\pi^2/g^2}$ in~\eqref{eq:Eth2}),
and $B_2$, $B_3$, \textit{etc.} are the 2-particle, 3-particle, and higher-particle 
interaction coefficients that encode systematic corrections to the ideal gas approximation.

Following the same logic, the expression~\eqref{eq:Eth2} for the vacuum energy $E(\theta)$
will receive power corrections of the form  
$\sum_{k=2}^\infty B_k \left(|K|\, e^{-8\pi^2/g^2}\right)^k$
multiplied by trigonometric functions of integer multiples of $\bar\theta$.  
For each $k$ there is a $k$-cluster of interacting instanton and anti-instanton 
configurations; each such cluster corresponds to a connected diagram.
\end{enumerate}

\paragraph{Key point:}
Using the \emph{dilute} instanton gas formalism to compute the partition function -- and, more generally, any correlator -- automatically generates the full set of \emph{disconnected} instanton diagrams.
The physically relevant \emph{connected} part of this set consists solely of the \emph{single}-instanton ($1$) and \emph{single}-anti-instanton ($\bar 1$) contributions,\footnote{%
This statement applies specifically to the DIGA framework, where the instanton gas is taken to be ideal, i.e.\ all interaction terms $B_{k\ge 2}$ are set to zero by construction.}
\begin{eqnarray}
\label{eq:E1p2}
E &=&E_{1} +E_{\bar 1} \,=\, \frac{- \, \log Z_{\rm DIGA}}{V_4},\\ 
\label{eq:Gconn}
\langle {\rm Correlator}\rangle_{\rm conn}  &=& \langle {\rm Correlator}\rangle_{1} +\langle {\rm Correlator}\rangle_{\bar 1} 
\,=\,
\frac{\langle {\rm Correlator}\rangle_{\rm DIGA}}{Z_{\rm DIGA}}.
\end{eqnarray}
The single-instanton contributions, $E_{1}(\theta)$ and
$\langle {\rm Correlator} \rangle_{1}(\theta)$, are entirely sufficient to determine the full $\theta$-dependence of physical observables.\footnote{On the right-hand side of~\eqref{eq:Gconn} one may also include the $0$-instanton (perturbative) contribution, $\langle{\rm Correlator}\rangle_{0}$, if it is non-zero. This does not modify the $\theta$-dependence.}
These quantities can be computed directly simply by evaluating the one-instanton contributions, without any reference to or need for performing the infinite summations over $n$ and $\bar n$ appearing in the disconnected DIGA expressions on the right-hand sides of~\eqref{eq:E1p2}-\eqref{eq:Gconn}.

\medskip

Let us now examine more closely the structure of a connected correlator in an instanton background. For a direct comparison with the ACGT approach, we consider the two-point correlation function of a massive quark in QCD (as in Eq.~(26) of Ref.~\cite{Ai:2024cnp}),
\begin{eqnarray}
\langle {\rm Correlator}\rangle_{\rm conn}  &\equiv&
\langle \psi(x_i) \bar\psi(x_j)\rangle_{\rm conn} \,=\, -\frac{\delta^2 \log Z[\eta,\bar\eta]}{\delta\bar\eta(x_i) \delta\eta(x_j)}|_{\bar\eta=0=\eta} \,=\,
\frac{\int DA_\mu D\bar\psi D\psi \,\psi(x_i) \bar\psi(x_j)\,e^{-S}}{\int DA_\mu D\bar\psi D\psi \,e^{-S}}
\end{eqnarray}
and compute the right hand side in the DIGA approximation.  We have 
\begin{eqnarray}
\label{eq:Gconn1}
\langle {\rm Correlator}\rangle_{\rm conn}  \,=\,
\frac{\langle {\rm Correlator}\rangle_{\rm DIGA}}{Z_{\rm DIGA}}\,=\,
\frac{\sum_{n=0}^\infty \sum_{\bar n=0}^\infty \, \langle {\rm Correlator}\rangle_{n,\bar n}}
{\sum_{n=0}^\infty \sum_{\bar n=0}^\infty \, Z_{n,\bar n}},
\end{eqnarray}
where $\langle {\rm Correlator}\rangle_{n,\bar n}$ and $Z_{n,\bar n}$ denote 
contributions of the ideal gas involving 
$n$-instantons and $\bar n$-anti-instantons. These are given as follows,
\begin{eqnarray}
\label{eq:Znnbar}
Z_{n,\bar n} &=&
\frac{(V_4 |E_1| e^{i\bar\theta})^n}{n!} \,\,
\frac{(V_4 |E_1| e^{-i\bar\theta})^{\bar n}}{\bar n!},\\
\label{eq:Gnnbar}
\langle {\rm Correlator}\rangle_{n,\bar n} &=&
\langle {\rm Correlator}\rangle_{1} \, Z_{n-1,\bar n} \,+\,
\langle {\rm Correlator}\rangle_{\bar 1} \, Z_{n,\bar n-1} \,+\,
\langle {\rm Correlator}\rangle_{0} \, Z_{n,\bar n},
\end{eqnarray}
where $\langle {\rm Correlator}\rangle_{1}$, $\langle {\rm Correlator}\rangle_{\bar 1} $ and
$\langle {\rm Correlator}\rangle_{0} $ denote  $\langle \psi(x_i) \bar\psi(x_j)\rangle$ computed in the background of a single instanton, single anti-instanton, and in perturbation theory respectively. 

 We see that each of the expressions in~\eqref{eq:Znnbar} and \eqref{eq:Gnnbar}
  corresponds to an $(n+\bar n)$-fold product of elementary 
single-instanton and single-anti-instanton contributions.
For example, the expression on the right-hand side of~\eqref{eq:Znnbar} is simply the product of $n$ single-instanton factors and $\bar n$ single-anti-instanton factors (divided by the appropriate Bose--Einstein symmetry factor).
Likewise, the first term on the right-hand side of~\eqref{eq:Gnnbar} represents the one-instanton contribution to the correlator multiplied by $(n-1)$ disconnected single-instanton contributions and by $\bar n$ single-anti-instanton contributions to the partition function.
All such terms are disconnected diagrams built from $(n+\bar n)$ instanton `blobs', exactly as expected in the dilute instanton gas picture.

After carrying out the sums over $n$ and $\bar n$ in both the numerator and the denominator in~\eqref{eq:Gconn1}, all of the $Z$-factors cancel, and we obtain the simple result,
\begin{eqnarray}
\label{eq:Gconn2}
\frac{\langle {\rm Correlator}\rangle_{\rm DIGA}}{Z_{\rm DIGA}} \,=\,
\langle {\rm Correlator}\rangle_{1} \,+\,
\langle {\rm Correlator}\rangle_{\bar 1} \,+\,
\langle {\rm Correlator}\rangle_{0},
\end{eqnarray}
which is the correct expression for the connected correlator $\langle {\rm Correlator}\rangle_{\rm conn}$ and is fully consistent with~\eqref{eq:Gconn}. 

\paragraph{Comments:} 
\begin{enumerate}

\item[{\bf 1.}] The correlators in~\eqref{eq:Gconn2} do depend on $\theta$ and this dependence can be made explicit,
\begin{eqnarray}
\label{eq:Gconn3}
\frac{\langle {\rm Correlator}\rangle_{\rm DIGA}}{Z_{\rm DIGA}} \,=\,
e^{i \bar \theta}\,\langle {\rm Correlator}\rangle_{1, \,\bar \theta=0} \,+\,
e^{-i \bar \theta} \, \langle {\rm Correlator}\rangle_{\bar 1, \,\bar \theta=0} \,+\,
\langle {\rm Correlator}\rangle_{\rm pert. th.} 
\end{eqnarray}

\item[{\bf 2.}]  At no stage in deriving \eqref{eq:Gconn2} we had to take an infinite volume limit. 
\item[{\bf 3.}]  The use of the dilute-instanton-gas formulae and the associated sums over instanton numbers is entirely unnecessary for evaluating $\langle {\rm Correlator}\rangle_{\rm conn}$.
This quantity can be computed directly in the single-instanton approximation (with perturbative contributions added if desired), and given by the right-hand side of~\eqref{eq:Gconn2}.
%\begin{eqnarray}
%\label{eq:Gconn4}
 %\langle {\rm Correlator}\rangle_{\rm conn} \,=\,
%e^{i \bar \theta}\,\langle {\rm Correlator}\rangle_{1, \,\bar \theta=0} \,+\,
%e^{-i \bar \theta} \, \langle {\rm Correlator}\rangle_{\bar 1, \,\bar \theta=0} \,+\,
%\langle {\rm Correlator}\rangle_{\rm pert. th.} 
%\end{eqnarray}
Note that it is always the connected part of the correlator that is directly related to terms in the effective action.
\item[{\bf 4.}] When interactions are included, additional genuine multi-instanton
contributions can appear in~\eqref{eq:Gconn2}.  
These correspond to true multi-instanton ADHM solutions, whose effects have been
explicitly computed in $N=2$ and $N=4$ SYM theories in
Refs.~\cite{Finnell:1995dr,Dorey:1996hu,Nekrasov:2002qd,Dorey:1999pd}.  
Such multi-instanton contributions are unrelated to the dilute gas
approximation; rather, they represent systematic $k$-instanton corrections in
the sense of the virial expansion~\eqref{eq:virial}.
These effects do not play a role in our present analysis of $\theta$-dependence
to leading order in the instanton number.
 
\end{enumerate}

\medskip

Now we are ready to compare the outlined above standard calculation of $\theta$-dependence to an alternative
approach advocated in Refs.~\cite{Ai:2020ptm,Ai:2024cnp}.

\paragraph{The `absence of strong CP' approach:} The authors of~\cite{Ai:2020ptm,Ai:2024cnp} proposed to re-organise the 
infinite two-fold summation in the DIGA formula~\eqref{eq:Gconn1} into two steps, and apply the infinite volume limit after completing the first sum but before the second. 

Specifically, the sum over $n$ and $\bar n$ 
is re-arranged by first holding the net instanton number $\Delta n = n-\bar n$ fixed while summing over individual values of $n$, and second, summing over all values of $\Delta n$. In so far this procedure is
 harmless\footnote{The radius of convergence of the exponent and also of the Bessel functions appearing below is infinite, so the convergence criterium for the sums is $V_4 e^{-8\pi^2/g^2} |K| < \infty$ which works for any arbitrarily large finite volume}. Equation~\eqref{eq:Gconn1}
can be presented in an equivalent form as,
\begin{eqnarray}
\label{eq:GconnDn}
\langle {\rm Correlator}\rangle_{\rm conn}  \,=\,
\frac{\langle {\rm Correlator}\rangle_{\rm DIGA}}{Z_{\rm DIGA}}\,=\,\,
\lim_{N\to \infty}
\frac{\sum_{\Delta n=-N}^N \sum_{n}\, \langle {\rm Correlator}\rangle_{n,\,\bar n=n-\Delta n}
}{
\sum_{\Delta n=-N}^N \sum_{n} \, Z_{n,\,\bar n=n-\Delta n}
}.
\end{eqnarray}
Performing the inner summations over $n$ at fixed $\Delta n$ in the numerator and the denominator
in~\eqref{eq:GconnDn} 
we find,
\begin{eqnarray}
\nonumber
\sum_{n}\, &&\langle {\rm Correlator}\rangle_{n,\,\bar n=n-\Delta n} \,=\, \\
&&
\label{eq:num}
e^{i \Delta n \bar \theta}
\left(
I_{\Delta n-1}(x) \, \langle {\rm Correlator}\rangle_{1, \,\bar \theta=0} \,+\, 
I_{\Delta n+1}(x) \, \langle {\rm Correlator}\rangle_{\bar 1, \,\bar \theta=0} \,+\, 
I_{\Delta n}(x) \, \langle {\rm Correlator}\rangle_{0} 
\right),
\end{eqnarray}
and
\begin{eqnarray}
\label{eq:den}
\sum_{n} \, Z_{n,\,\bar n=n-\Delta n} \,=\,
e^{i \Delta n \bar \theta}
\,
I_{\Delta n}(x).
\end{eqnarray}
 Here 
$I_{\Delta n\pm1}(x)$ and $I_{\Delta n}(x)$ are the modified Bessel functions of the first kind, and their argument is
\[x= 2 V_4 \, e^{-8\pi^2/g^2} |K|.\]
If we sum these expressions in~\eqref{eq:num}-\eqref{eq:den} over $\Delta n$ and evaluate the ratio on the right-hand side of~\eqref{eq:GconnDn}, we recover the result~\eqref{eq:Gconn2} for the connected correlator $\langle {\rm Correlator}\rangle_{\rm conn}$.
It is important to stress, that there are no infinities appearing in the final expression on the right hand side of~\eqref{eq:Gconn2} associated with an infinite volume. The integration over the instanton  position $x_0$ is soaked up by the $(x_i-x_0)$ and $(x_i-x_0)$ arguments of the two fermion zero modes in the correlator 
$\langle \psi(x_i) \bar\psi(x_j)\rangle_{1}$ (or for the anti-instanton, in $\langle \psi(x_i) \bar\psi(x_j) \rangle_{\bar 1}$) and no explicit factors of $V_4$ are left in~\eqref{eq:Gconn2}. So, in the standard statistical mechanics-inspired ideal instanton gas approach, one keeps the volume finite: as already noted, the sums in~\eqref{eq:Zth2} and~\eqref{eq:Gconn1} converge for any finite value of $x$, and  the dependence on the spacetime volume cancels in the ratio appearing in~\eqref{eq:Gconn2}. 
 
 \medskip
 
 The approach advocated in Refs.~\cite{Ai:2020ptm,Ai:2024cnp} is different. These authors insist on taking 
 the infinite volume limit before  the infinite sum over $\Delta n$. 
 Specifically, they define the correlator,
 \begin{eqnarray}
\label{eq:G_AGT}
\langle {\rm Correlator}\rangle_{\rm ACGT}  \,=\,
%\frac{\langle {\rm Correlator}\rangle_{\rm DIGA}}{Z_{\rm DIGA}}\,=\,\,
\lim_{N\to \infty}\, \lim_{V_4 \to \infty}
\frac{\sum_{\Delta n=-N}^N \sum_{n}\, \langle {\rm Correlator}\rangle_{n,\,\bar n=n-\Delta n}
}{
\sum_{\Delta n=-N}^N \sum_{n} \, Z_{n,\bar n=n-\Delta n}
}
\end{eqnarray} 
imposing the $\lim_{V_4 \to \infty}$
 \emph{prior} to extending the summation over $\Delta n$ to an infinite domain.
 In these settings, one makes use of the asymptotic values of the Bessel functions,
\begin{equation}
I_m (x) \,\approx\, \frac{1}{\sqrt{2\pi x}}\,\,  e^x\,,    \quad 
{\rm as} \,\, x\to \infty,    \quad 
{\rm for \,\, any} \,\, m\in Z,
\end{equation}
so that all ratios of $I_{\Delta n\pm1}(x)/ I_{m}(x)$ and $I_{\Delta n}(x)/ I_{m}(x)$ are universal and $=1$ in the $V_4\to \infty$ limit.
Normalising the numerator and denominator expressions~\eqref{eq:num}-\eqref{eq:den} by the same 
$I_m(x)$ and taking the limit $x\to \infty$, one obtains for the numerator of~\eqref{eq:G_AGT}:
\begin{eqnarray}
\label{eq:num2}
\frac{1}{\sqrt{2\pi x}}\,\,  e^x
\left(
\langle {\rm Correlator}\rangle_{1, \,\bar \theta=0} \,+\, 
\langle {\rm Correlator}\rangle_{\bar 1, \,\bar \theta=0} \,+\, 
\langle {\rm Correlator}\rangle_{0} 
\right)
\left(\sum_{\Delta n=-N}^N
e^{i \Delta n \bar \theta}\right),
\end{eqnarray}
and for the denominator:
\begin{eqnarray}
\label{eq:den2}
\frac{1}{\sqrt{2\pi x}}\,\,  e^x
\left(\sum_{\Delta n=-N}^N
e^{i \Delta n \bar \theta}\right),
\end{eqnarray}
Taking the ratio of these expressions, one immediately sees that the entire $\theta$-dependence cancels out in~\eqref{eq:G_AGT},
\begin{eqnarray}
\label{eq:G_AGT2}
\langle {\rm Correlator}\rangle_{\rm ACGT}  \,=\,
\langle {\rm Correlator}\rangle_{1, \,\bar \theta=0} \,+\,
 \langle {\rm Correlator}\rangle_{\bar 1, \,\bar \theta=0} \,+\,
\langle {\rm Correlator}\rangle_{\rm pert. th.} 
\end{eqnarray}  
The authors of Refs.~\cite{Ai:2020ptm,Ai:2024cnp} take this cancellation to imply that physical observables, as defined by the order of limits in~\eqref{eq:G_AGT}, are insensitive to the CP-violating parameter $\theta$. This assertion forms the core of their claim that CP violation is absent in QCD.

\medskip

I will now argue that taking the infinite volume limit before completing the second infinite sum (i.e before
taking $N\to \infty$ in~\eqref{eq:G_AGT}) invalidates the structure of the numerator and denominator in this expression as sums over disconnected diagrams.

Let us examine the denominator first. Equation~\eqref{eq:den2} shows that the partition function takes the form,
\begin{equation}
Z\,=\, \lim_{N\to \infty}\, \lim_{V_4 \to \infty}
\sum_{\Delta n=-N}^N \sum_{n} \, Z_{n,\bar n}\,=\,
\lim_{N\to \infty}\, 
\left(\sum_{\Delta n=-N}^N
e^{i \Delta n \bar \theta}\right)\,
\lim_{x \to \infty} \,\frac{1}{\sqrt{2\pi x}}\,\,  e^x
\end{equation}
To simplify matters further we can temporarily set $\bar\theta=0$, so that,
\begin{equation}
Z\,=\, 
\left[\lim_{N\to \infty} (2N+1) \right] \times
\left[\lim_{x \to \infty} \,\frac{1}{\sqrt{2\pi x}}\,\,  e^x\right]
\end{equation}
Only the factor $e^x=e^{\, 2 V_4  |K| e^{-8\pi^2/g^2}}$ in this expression can be interpreted as the 
sum of disconnected instanton and anti-instanton contributions, in accordance with the result~\eqref{eq:Zth2} evaluated at $\bar\theta=0$. However, this interpretation is spoiled by the additional multiplicative factor
\[
\frac{1}{\sqrt{2\pi x}}\,=\, \frac{1}{\sqrt{2\pi}}\, \frac{e^{4\pi^2/g^2}}{\sqrt{2 V_4  |K|}},
\]
which introduces a square-root dependence on the instanton density together with an explicit inverse power of the (infinite) spacetime volume.
This is further compounded by the overall factor $2N$ which itself diverges in the limit of $N\to \infty$. 

Applying the same line of reasoning to the numerator, one finds that the expression~\eqref{eq:num2} in~\eqref{eq:G_AGT} similarly lacks the structure of a sum over disconnected contributions to the correlator.

\medskip
\medskip

The bottom line is that the object formally defined in~\eqref{eq:G_AGT} and computed in~\eqref{eq:G_AGT2} does not appear to possess any clear physical meaning or interpretation.
\begin{enumerate}
\item[{\bf 1.}] Since the framework of Refs.~\cite{Ai:2020ptm,Ai:2024cnp} does not attempt to go beyond the ideal, non-interacting instanton gas and does not incorporate genuine \emph{multi}-instanton contributions, it is particularly puzzling that the instanton correlators arising in their construction differ from the leading, single-instanton contributions on the right-hand side of~\eqref{eq:Gconn2}.
\item[{\bf 2.}] The role normally played by dressing the minimal single-instanton correlators with the dilute instanton gas is to generate all disconnected diagrams. Once the connected graphs are extracted from this enlarged set, one should recover precisely the original single-instanton correlators. The order-of-limits prescription imposed in~\eqref{eq:G_AGT} obscures this standard construction. Consequently, after taking these limits, the quantity
$\langle {\rm Correlator}\rangle_{\rm ACGT}$ can no longer be interpreted as the proper connected part of the correlator within the DIGA approximation presented in~\eqref{eq:Gconn1}.
\item[{\bf 3.}] The rationale in Refs.~\cite{Ai:2020ptm,Ai:2024cnp} for taking the large-$V_4$ limit before summing over all topological sectors~$\Delta n$ was to localise instantons well inside $V_4$, sufficiently far from the boundary at infinity. This was intended to ensure that, in every topological sector, the gauge fields approach the required pure-gauge boundary conditions. 

However, such concerns can be safely bypassed by expanding path integral around BPST instantons in \emph{singular} gauge. In this gauge, the instanton gauge fields fall off as $\rho^2/x^3$ and therefore approach zero -- not a non-trivial pure gauge -- at large~$|x|$. The topological winding is instead encoded around the (pure-gauge) singularity at the instanton centre. The singularity itself is harmless, being a removable gauge artefact. Indeed, it is well known that instantons in singular (rather than regular) gauge are the correct configurations to use when computing, for example, scattering amplitudes: their rapid fall-off guarantees that the LSZ reduction of the instanton background is non-vanishing on-shell, 
\begin{equation}
 A^{a\, {\rm inst}}_{LSZ}(p,\lambda)\,=\, \lim_{p^2\to 0} p^2 \epsilon^\mu(p,\lambda) \, A_\mu^{a\, {\rm inst}}(p)
 \,=\, \epsilon^\mu(p,\lambda) \,\bar{\eta}^a_{\mu\nu} p_\nu \, \frac{4i\pi^2 \rho^2}{g} \, e^{ip \cdot x_0}\,
 \neq 0
 \,. \label{eq:LSZ}
\end{equation}
while regular-gauge instantons would give a trivially vanishing contribution.

In the presence of scalar field VEVs the gauge fields are massive and in practice one employs the so-called `constrained' instantons~\cite{Affleck:1980mp}
whose gauge fields decay as $A_\mu \sim e^{-m|x|}$ at large distances outside the instanton core. In this case instantons are always automatically localised within the spacetime volume $V_4$: the gauge fields vanish at infinity rather than approaching a pure-gauge configuration. Consequently, there is no valid justification for taking the infinite-volume limit prior to summing over distinct instanton-charge sectors.
 
Overall, I see no compelling reason to impose the infinite-volume limit in~\eqref{eq:G_AGT} in the first place.

\item[{\bf 4.}] {\tt In summary}, we conclude that one should simply employ standard instanton calculus and compute the instanton contributions to the relevant connected correlators directly, as in~\eqref{eq:Gconn}:
\begin{eqnarray}
\label{eq:Gconn222}
\langle {\rm Correlator}\rangle_{\rm conn}  \,=\, \langle {\rm Correlator}\rangle_{1} +\langle {\rm Correlator}\rangle_{\bar 1} \,+\, \ldots
\end{eqnarray}
where the ellipsis denotes higher-order contributions from connected multi-instanton clusters. The correlators on the right-hand side exhibit the standard dependence on the $\theta$-parameter, as shown in~\eqref{eq:Gconn3}, with $\langle {\rm Correlator}\rangle_{1}\propto e^{i\bar\theta}$ and
$\langle {\rm Correlator}\rangle_{\bar 1}\propto e^{-i\bar\theta}$.

\bigskip

\centerline{*******}

\item[{\bf 5.}] {\tt Theta-dependence beyond QCD}:  Instanton contributions in 4-dimensional gauge theories always carry complex phases
which are naturally organised as a Fourier series in
$q = e^{2\pi i\tau}$, where $\tau = \frac{\theta}{2\pi} + \frac{4\pi i}{g^2}$. In the Seiberg--Witten $\mathcal N=2$ supersymmetric $SU(2)$ gauge theory~\cite{Seiberg:1994rs}, the low-energy effective action is encoded in the holomorphic prepotential 
$\mathcal F(a,\tau)$ where $a$ is the VEV of the $\mathcal N=1$ chiral superfield $\Phi$.
For large $|a|\gg |\Lambda_{\sst\mathcal N=2}|$, the theory is weakly coupled and one obtains an expansion
\begin{eqnarray}
\label{eq:prep1}
\mathcal F(a,\tau)
&=& \mathcal F_{\mathrm{\rm pert}}(a,\tau)
+ \sum_{k\ge 1} \mathcal F_k(a)\, e^{2\pi i k \tau}\\
&=&
\label{eq:prep2}
{i\over2\pi}
a^2\,\log{2a^2\over e^3\Lambda_{\sst\mathcal N=2}^2}\ -\ {i\over\pi}
\sum_{k=1}^\infty
F_k \,\left({\Lambda_{\sst\mathcal N=2}\over a}\right)^{4k}a^2
\end{eqnarray}
where the summation is over all instanton numbers $k$. 
Multi-instanton coefficients \(F_k\) were computed from first principles in
Refs.~\cite{Finnell:1995dr,Dorey:1996hu,Nekrasov:2002qd} and were shown to agree
precisely with the Seiberg--Witten predictions~\cite{Seiberg:1994rs}. The complex
phases \(e^{2\pi i k \tau}\) of the multi-instanton contributions appearing in
line~\eqref{eq:prep1} are essential for ensuring holomorphy and for reproducing
the correct monodromy structure of the prepotential.
The complex-valued dimensional transmutation scale
\(\Lambda_{\sst\mathcal N=2}\) in~\eqref{eq:prep2} is given by the standard
RG-invariant (in the supersymmetric setting) expression
\begin{equation}
\label{eq:LamN2}
\Lambda_{\sst\mathcal N=2}^{b_0}
\,=\, \mu^{b_0}\, e^{2\pi i \tau(\mu)}
\,=\, \mu^{b_0}\, e^{-\frac{8\pi^2}{g^2(\mu)} + i \theta}
\,, \qquad b_0 = 2N_c = 4 \, ,
\end{equation}
where \(\mu \in \mathbb{R}\) is the RG scale, which may be chosen as
\(\mu = |a|\). The dependence on the \(\theta\)-parameter in
line~\eqref{eq:prep2} therefore remains explicit and is encoded in the phase of
\(\Lambda_{\sst\mathcal N=2}\).
This phase can be combined with the phase of the scalar VEV \(a\)
into a single physical CP-odd parameter,
\begin{equation}
\bar\theta \;=\; \theta - 4\,\arg a \, ,
\end{equation}
which is invariant under anomalous \(U(1)_R\) transformations.

Turning now to the case of $\mathcal N=4$ SYM theory,
we note that the factors of $e^{2\pi i k \tau}$
allow non-perturbative correlators  to be organised into genuine modular or quasi-modular forms~\cite{Banks:1998nr,Green:1997tv} and \cite{Dorey:1999pd}.
They implement S-duality by providing the correct transformation behaviour under $SL(2,\mathbb Z)$.

We see that $\theta$-phases are not a fantom, instead they play an essential role in the non-perturbative QFT dynamics based on well-established exact results in supersymmetric YM models.

\item[{\bf 6.}] {\tt Strong CP from SUSY}: To actually generate a $\bar\theta$-dependent vacuum energy, $E(\bar\theta)$, aka the axion potential, we have to softly break supersymmetry.\footnote{Otherwise the vacuum energy is trivially vanishing in any supersymmetric vacuum due to an exact cancellation between bosonic and fermionic contributions.} 
Arguably, the simplest starting point is an $\mathcal N=1$ supersymmetric $SU(2)$ pure gauge theory.
The theory has a mass gap and there are two (confining) supersymmetric vacua, 
$|{\rm vac}\rangle_{\pm}$ characterised by the corresponding values of the gluino condensate,
\begin{equation}
\label{eq:lam22}
\frac{1}{16 \pi^2}\, \langle {\rm tr} \lambda^2 \rangle_{\pm} \,=\, \pm \, \Lambda_{\sst \mathcal{N}=1}^3
\,=\, \pm \,  e^{i \frac{1}{2} \theta}\, |\Lambda_{\sst \mathcal{N}=1}|^3.
\end{equation}
where the complex-valued dimensional transmutation scale of the $\mathcal N=1$ theory is 
({\it cf.} eq.~\eqref{eq:LamN2}),
 \begin{equation}
\label{eq:LamN1}
\Lambda_{\sst \mathcal{N}=1}^{3}
\,=\, \mu^{3}\, e^{\, \frac{1}{2}2\pi i \tau(\mu)}\,=\, 
\mu^{3}\, e^{-\frac{4\pi^2}{g^2(\mu)}\,+\, i \frac{1}{2}\theta}
\,, \quad b_0=3N_c=6
\end{equation}
Since $b_0=6$ in our model, the expression for $\Lambda_{\sst \mathcal{N}=1}^{3}$ in~\eqref{eq:LamN1} 
corresponds to a square root of a \mbox{1-instanton} factor, or more precisely, to a contribution from a fractional instanton carrying instanton charge $1/2$. 

This reasoning was made precise in Refs.~\cite{Davies:1999uw,Davies:2000nw} where the where the theory was compactified on $\mathbb{R}^3 \times S^1$. The gauge field component along the compact direction acquires a vacuum expectation value $\langle A_4^{(3)} \rangle \, = \, v ~\sim 1/R$, where $R$ is the radius of $S^1$. In the small radius regime, $R \ll 1/|\Lambda_{\sst \mathcal{N}=1}|$, the theory is weakly-coupled, 
with $v/|\Lambda_{\sst \mathcal{N}=1}| \gg 1$, and a semi-classical approximation is therefore justified. 
The semi-classical configurations of minimal Euclidean action on $\mathbb{R}^3 \times S^1$ are the monopole-instantons which carry the instanton charge  $1/2$. 
For gauge group $SU(2)$ there are two such configurations on $\mathbb{R}^3 \times S^1$, constructed in Refs.~\cite{Lee:1998bb,Kraan:1998pm}, commonly referred to as the BPS and KK monopoles. 
Their semi-classical contributions the gluino condensate were computed in~\cite{Davies:1999uw},
\begin{equation}
\VEV{\tr \lambda^2\over16\pi^2}_{\sst\rm BPS\,mono} \ =\ 
\VEV{\tr \lambda^2\over16\pi^2}_{\sst\rm KK\,mono}\ =\ 
\hf\,  \mu^3 \, e^{-{4\pi^2 \over g^2(\mu)}\,+\, i \frac{1}{2} \theta}\,=\,
\hf\, e^{i \frac{1}{2} \theta}\, |\Lambda_{\sst \mathcal{N}=1}|^3.
\label{lctw}
\end{equation}
Summing the contributions of both monopole configurations reproduces the value
of the gluino condensate in~\eqref{eq:lam22}. The relative sign of the condensate
in the two vacua is recovered by shifting the \(\theta\)-parameter in~\eqref{lctw}
by \(2\pi\). Importantly, since the gluino condensate is a holomorphic \(F\)-term,
its value, computed in the weakly coupled regime at small \(R\), can be
analytically continued to the decompactified, strongly coupled theory by taking
the limit \(R \to \infty\). In this limit, the result~\eqref{eq:lam22} remains
valid~\cite{Davies:1999uw}.

There is also an alternative and complementary semiclassical approach: the
weak-coupling instanton calculation of Novikov, Shifman, Vainshtein, and Zakharov~\cite{Novikov:1985ic}. 
In this framework, the gluino
condensate~\eqref{eq:lam22} is computed in a theory augmented by additional
matter fields, whose presence renders the model weakly coupled and justifies a
conventional constrained instanton analysis. Holomorphy of
supersymmetric \(F\)-terms is then exploited to decouple the extra matter fields
and to flow to the original pure \(\mathcal N=1\) gauge theory, once again
reproducing the expression on the right-hand side of~\eqref{eq:lam22}, including
its crucial \(\theta\)-dependence.

We now break supersymmetry to $\mathcal N=0$ by introducing a mass term for the gluino,
\begin{equation}
\label{eq:N=0}
\Delta\mathcal L_{\rm soft}
=  m_\lambda\,\lambda\lambda + \text{h.c.}
\end{equation}
The soft mass term above involves a complex-valued 
mass parameter $m_\lambda$, and the ordering, $ |m_\lambda| \ll \Lambda_{\sst\mathcal N=1}$, ensures that this is a small deformation of the original theory.
As a result, the vacuum energy \(E\), which vanishes in
the supersymmetric theory, is lifted by the soft breaking terms in
\eqref{eq:N=0}, yielding:
\begin{equation}
\label{eq:2branches}
\Delta E_{\pm}
\,=\, -\left(m_\lambda \langle \lambda\lambda\rangle_{\pm}
  \,+\,  m_\lambda^* \langle \bar\lambda\bar\lambda\rangle_{\pm}\right) 
   \,=\, \mp \,\varepsilon\, \cos\left(\bar\theta/2\right)
\end{equation}
to first order in $m_\lambda$. Here we defined 
$\varepsilon\,=\, 32\pi^2 |m_\lambda \Lambda_{\sst\mathcal N=1}|$ 
and introduced the physical $\bar\theta$ phase, which is invariant under chiral $U(1)_R$ rotations,
%\begin{equation}
$\bar\theta \,=\, \theta+2\,{\rm arg}\, m_\lambda$.
%\end{equation}

The vacuum energy density is obtained by taking the lower envelope of
the two branches,
\begin{equation}
\label{eq:2branchesC}
E(\bar\theta) \,=\, {\rm min}\, \left(-\, \varepsilon \cos (\bar\theta/2)\, ,\, +\,\varepsilon \cos (\bar\theta/2)\right)
\end{equation}
As \(\bar\theta\) is varied, a first-order phase transition occurs at
\(\bar\theta = \pi\), where the theory switches from one branch to the other.
\begin{figure}[t]
    \centering
   {{\includegraphics[width=6cm]{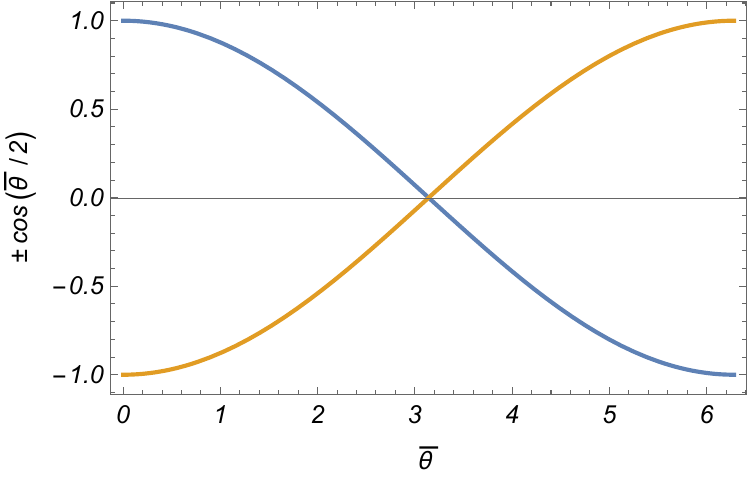} }}%
    \qquad \quad
{{\includegraphics[width=6cm]{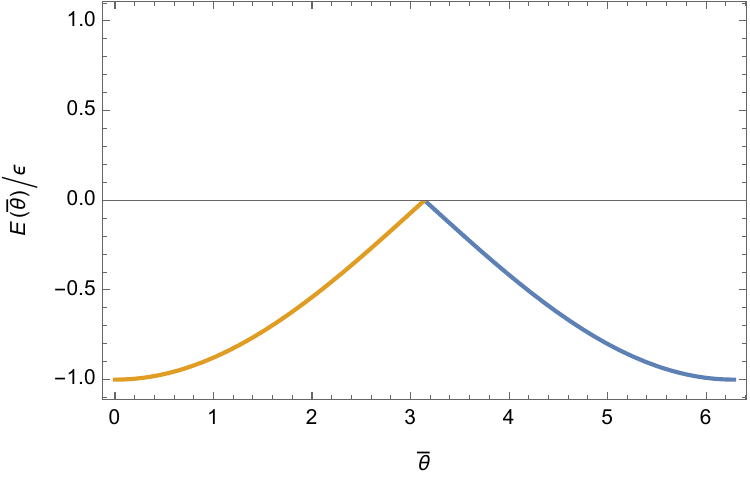} }}%
    \caption{The figure on the left displays the two branches of the vacuum energy in~\eqref{eq:2branches}.
    The physical (combined multi-branch) vacuum energy density $E(\bar\theta)$ of the $SU(2)$ YM theory of~\eqref{eq:2branchesC} is shown on the right. $E(\bar\theta)$ is a periodic function with a period $2\pi$ and cusps at $\bar\theta=\pi$ (mod $2\pi$). }%
    \label{fig:example}%
\end{figure}
The resulting multi-branch structure $E(\bar\theta) $, or equivalently of the axion potential, \eqref{eq:2branchesC}
 is plottted 
in the right-hand-side panel of Fig.~\ref{fig:example}. 
Our simple \(SU(2)\) Yang--Mills model thus provides a concrete realisation of
Witten's general multi-branch periodic potential, originally motivated in the
large-\(N_c\) limit by Witten, Di Vecchia, and Veneziano in
Refs.~\cite{Witten:1978bc,DiVecchia:1980yfw}.

\item[{\bf 7.}] {\tt Strong CP in the softly broken Seiberg-Witten theory}: It is also well-known that the same
non-trivial (multi-brach) periodic structure of $E(\bar\theta)$ follows from the 
$\mathcal N=2$ super--Yang--Mills theory broken to $\mathcal N=0$ either by introducing soft gluino mass terms
\cite{Konishi:1996iz}
or by turning on an anomaly mediated supersymmetry breaking~\cite{Csaki:2025fkh}
The Coulomb branch of the Seiberg-Witten model is parametrised by the flat direction
\begin{equation}
u \equiv \langle {\rm tr} \Phi^2 \rangle,
\end{equation}
where $\Phi$ is the $\mathcal N=1$ chiral superfield. 
SUSY breaking $\mathcal N=2$ to $\mathcal N=1$ is achieved by adding the mass-term for $\Phi$,
\begin{equation}
\label{eq:N=1}
W_{\mathcal N=1} = m\,{\rm tr}\Phi^2,
\end{equation}
in which case the flat direction is lifted producing two isolated $\mathcal N=1$ vacua,
located near the Seiberg--Witten singularities
\begin{equation}
\label{eq:SWvac}
u_{\pm}\,=\, \langle {\rm tr} \Phi^2 \rangle_\pm \,=\, \pm\, \Lambda_{\sst\mathcal N=2}^2 
\end{equation}
Next step is to break $\mathcal N=1$ to  $\mathcal N=0$ 
by soft gluino masses, as in \eqref{eq:N=0}.
The vacuum energy of the resulting pure $SU(2)$ gauge theory is given by~\cite{Konishi:1996iz},
\begin{equation}
\label{eq:2branchesC2}
E(\bar\theta) \,=\, {\rm min}\, \left( -\,\varepsilon \cos (\bar\theta/2)\, ,\, +\, \varepsilon \cos (\bar\theta/2)\right),
\end{equation}
where the energy-density scale $\varepsilon$ is now
$\varepsilon\,=\, 32\pi^2 |m\,m_\lambda \Lambda_{\sst\mathcal N=2}|$ 
and the invariant CP-odd phase is
$\bar\theta \,=\, \theta+2\,({\rm arg}\, m_\lambda\,+\, {\rm arg}\, m)$.
The functional form of \(E(\bar\theta)\) in~\eqref{eq:2branchesC2} is
identical to that obtained above in~\eqref{eq:2branchesC} and plotted in
Fig.~\ref{fig:example}. The result is derived in the parametric regime
\(|m_\lambda| \ll |m| \ll |\Lambda_{\sst\mathcal N=2}|\), where the soft
breaking scales may be treated as small deformations of the underlying
Seiberg--Witten theory.

This construction also provides a natural explanation for the multi-branch
structure of the vacuum energy \(E(\bar\theta)\). The weakly coupled degrees of
freedom in the neighbourhood of the two vacua \(u_\pm\) are mutually non-local.
The branch \(-\,\varepsilon \cos(\bar\theta/2)\) is associated with the
\(u_+\) vacuum, in which the \((1,0)\) monopole condenses, whereas the second
branch, \(+\,\varepsilon \cos(\bar\theta/2)\), corresponds to the \(u_-\)
vacuum, where the condensing field is the \((1,1)\) dyon.
For \(-\pi < \bar\theta < \pi\), the appropriate weakly coupled degrees of
freedom are monopoles, while in the interval \(\pi < \bar\theta < 3\pi\) the
relevant description involves dyons.

In a subsequent study, Ref.~\cite{Csaki:2025fkh} analysed the $\theta$-dependence of $\mathcal N=2$ $SU(2)$ QCD with $N_f$ fundamental flavours, explicitly broken to $\mathcal N=1$ and coupled to anomaly-mediated supersymmetry breaking. For $N_f=0$, this analysis reproduces the potential shown in Fig.~\ref{fig:example}, and generalises it to $N_f=1,2,3$, revealing cusps and associated phase transitions in $E(\bar\theta)$ at fractional values of $\bar\theta$.

\item[{\bf 8.}] {\tt Conclusion -- when instantons are and are not enough:} 
Instanton-based semiclassical methods are reliable only in weakly coupled
theories. QCD in normal conditions does not
satisfy this criterion. Consequently, the use of instantons to determine the
\(\theta\)-dependence of physical observables requires modifying the theoretical
setting so as to access a weakly coupled phase. One may then apply
instanton-based techniques and subsequently assess whether the resulting
insights can be meaningfully extrapolated to the original strongly coupled
theory.
A phenomenologically relevant example is provided by QCD at high temperature,
\(T \gg \Lambda\). In this regime the theory becomes weakly coupled, the
instanton formalism~\cite{Gross:1980br} is well justified, and the vacuum energy
\(E(\bar\theta; T)_{\rm QCD}\) can be reliably computed using the dilute
instanton gas approximation (DIGA).

Another class of controlled modifications involves introducing scalar fields
with vacuum expectation values much larger than the strong-coupling scale,
\(\langle\phi\rangle \gg \Lambda\). In such settings (which include applications
to the electroweak theory and to supersymmetric models) one employs constrained instanton configurations~\cite{Affleck:1980mp}.
These can be used to compute, for example, the Seiberg--Witten
prepotential~\cite{Finnell:1995dr,Dorey:1996hu,Nekrasov:2002qd}
or the gluino condensate~\cite{Novikov:1985ic}. The resulting expressions can
then be analytically continued to the strongly coupled regime, as discussed
above, thereby recovering the  \(\theta\)-dependence of the gauge theory.

One can also consider compactifying compactifying spacetime on
\(\mathbb{R}^3 \times S^1\), which allows the use of fractionally charged
instanton--monopole configurations~\cite{Davies:1999uw}. Other more recent
applications of semiclassical techniques to the computation of the
\(\theta\)-dependence and axion potentials based on small instanton-like
configurations in gauge theories, can be found in
Refs.~\cite{Agrawal:2017ksf,Csaki:2019vte,Gherghetta:2020keg}.

\medskip

In all of the scenarios considered above, the $\theta$-dependence is non-trivial. 

\end{enumerate}

\medskip

\noindent In this note, I have first focused on instanton-based derivations of the
\(\theta\)-dependence in theoretical settings where the gauge dynamics is
weakly coupled, and subsequently on examples involving softly broken
supersymmetric theories at strong coupling. For a selection of complementary
formulations and alternative approaches, the reader is referred to the ongoing
discussion in the recent literature, including
Refs.~\cite{Benabou:2025viy,Ai:2025quf,Ringwald:2026apz}.

{\section*{Acknowledgements}}
I thank Steve Abel, Rodrigo Alonso, Nick Dorey, Andreas Ringwald and Michael Spannowsky for comments and useful discussions. \\

\medskip

\bibliographystyle{utphys}
\bibliography{biblio}{}
  
\end{document}